\documentclass[preprint,showpacs,aps]{revtex4}
\usepackage{graphicx}
\usepackage{epstopdf}
\usepackage{amsmath}
\usepackage{bm}
\usepackage{multirow}

\begin{document}

\title{Electronic structures of ternary iron arsenides AFe$_2$As$_2$ (A=Ba, Ca, or Sr)}
\author{Fengjie Ma$^{1,2}$}
\author{Zhong-Yi Lu$^{1}$}\email{zlu@ruc.edu.cn}
\author{Tao Xiang$^{3,2}$}\email{txiang@aphy.iphy.ac.cn}

\date{August 14, 2009}

\affiliation{$^{1}$Department of Physics, Renmin University of
China, Beijing 100872, China}

\affiliation{$^{2}$Institute of Theoretical Physics, Chinese
Academy of Sciences, Beijing 100190, China }

\affiliation{$^{3}$Institute of Physics, Chinese Academy of
Sciences, Beijing 100190, China }

\begin{abstract}

We have studied the electronic and magnetic structures of the
ternary iron arsenides AFe$_2$As$_2$ (A = Ba, Ca, or Sr) using the
first-principles density functional theory. The ground states of
these compounds are in a collinear antiferromagnetic order,
resulting from the interplay between the nearest and the
next-nearest neighbor superexchange antiferromagnetic interactions
bridged by As $4p$ orbitals. The correction from the spin-orbit
interaction to the electronic band structure is given. The pressure
can reduce dramatically the magnetic moment and diminish the
collinear antiferromagnetic order. Based on the calculations, we
propose that the low energy dynamics of these materials is described
effectively by a $t-J_H-J_1-J_2$-type model \cite{mafj}.

\end{abstract}


\maketitle

\section{Introduction}

The recent discovery of superconductivity in LaFeAsO by partial
substitution of O with F atoms below 26K\cite{kamihara} has
stimulated great interest on the investigation of physical
properties of iron-based pnictides. This type of quaternary
compounds consists of alternative tetrahedral FeAs and LaO layers
along the c-axis. The LaO layers act mainly as a charge reservoir.
The superconducting pairing occurs in the FeAs layers. More
recently, it was reported that the ternary iron-based arsenides
AFe$_2$As$_2$ (A=Ba, Ca, or Sr) become superconducting upon hole or
electron doping\cite{rotter2,sasmal,boyer,wu}. Similar as in
LaFeAsO\cite{cruz,mcguire}, these ternary iron arsenides also
exhibit a spin-density-wave-like anomaly and a structural transition
from the tetragonal $I4/mmm$ to the orthorhombic $Fmmm$ group at
some temperature between 140 K and 200 K \cite{rotter,bao}.
Furthermore, it was found that the high pressures can drive these
undoped ternary iron arsenides superconducting
\cite{torikachvili,park,alireza}.

To investigate the mechanism of superconductivity in these
materials, it is commonly believed that one needs to understand
first the electronic and magnetic structures of the parent
compounds. It has been shown that there is an essential similarity
of electronic states nearby the Fermi level in AFe$_2$As$_2$ (A=Ba,
Ca, or Sr) and LaFeAsO by studying the nonmagnetic
state\cite{mafj,nekrasov}. Moreover, the density of states at the
Fermi energy is only weakly doping dependent and the main effect of
doping is a change in the relative sizes of the electron and hole
Fermi surfaces\cite{singh2}.

In this paper, we report the electronic structures and magnetic
orders and properties of AFe$_2$As$_2$ (A=Ba, Ca, or Sr) obtained
from the first-principles electronic structure calculations. By
comparison of the energy of the non-magnetic state with those of a
number of magnetic ordered states, we find that, similar as in
LaFeAsO, the ground state of AFe$_2$As$_2$ is in a collinear
antiferromagnetic order. We have also studied the spin-orbit
interaction and the pressure effect in these materials. The
electronic and magnetic structure is found to be strongly affected
by the pressure effect, but weakly by the spin-orbit interaction.

\section{Computational Approach}

\begin{figure}
\includegraphics[width=8cm,height=4cm]{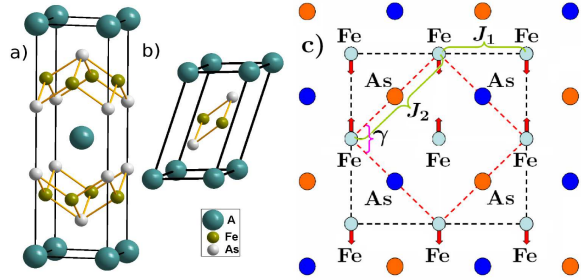}
\caption{(Color online) AFe$_2$As$_2$ (A=Ba, Ca, or Sr): a) a
tetragonal unit cell containing two formula units; b) a primitive
unit cell containing one formula unit; c) schematic top view of the
FeAs layer. The small red dashed square is an $a\times a$ unit cell,
while the large black dashed square is a $\sqrt{2}a\times \sqrt{2}
a$ unit cell. The Fe spins in the collinear antiferromagnetic order
are shown by arrows.} \label{conf}
\end{figure}

AFe$_2$As$_2$ (A=Ba, Ca, or Sr) takes the ThCr$_{2}$Si$_{2}$ type
structure and an ``A'' layer plays a similar role as a LaO layer in
LaFeAsO. Although AFe$_2$As$_2$ can be considered as a tetragonal
crystal with two formula units included in the corresponding unit
cell as shown in Fig.\ref{conf}(a), its primitive unit cell is
constructed by considering AFe$_2$As$_2$ as a triclinic crystal, in
which only one formula unit cell is included as shown in Fig.
\ref{conf}(b). In the calculations, we adopted the primitive cell as
calculation cell with the experimental lattice constants as the
input parameters. In the calculation of the electronic structures of
the nonmagnetic, the ferromagnetic and the square antiferromagnetic
Neel states, the $a\times a$ FeAs cell is taken as the base cell
(shown in Fig. \ref{conf}(c)). In the calculation of the collinear
antiferromagnetic state, the unit cell is doubled and the base cell
is the $\sqrt{2}a\times\sqrt{2}a$ FeAs cell as shown in Fig.
\ref{conf}(c).

In our calculations the plane wave basis method was used
\cite{pwscf}. We used the generalized gradient approximation (GGA)
of Perdew-Burke-Ernzerhof \cite{pbe} for the exchange-correlation
potentials. The ultrasoft pseudopotentials \cite{vanderbilt} were
used to model the electron-ion interactions. After the full
convergence test, the kinetic energy cut-off and the charge density
cut-off of the plane wave basis were chosen to be 600eV and 4800eV,
respectively. The Gaussian broadening technique was used and a mesh
of $16\times 16\times 8$ k-points were sampled for the
Brillouin-zone integration. The internal atomic coordinates within a
cell were determined by the energy minimization.

\section{Results and analysis}
\subsection{Nonmagnetic State}

We first studied the nonmagnetic state of the compound AFe$_2$As$_2$
(A = Ba, Sr, or Ca), which is the high temperature phase of these
materials. The electronic band structure of this state also provides
a reference for studying of the low temperature magnetic phases.
This can help us to understand the mechanism or the interactions
that drive the magnetic phase transition and the related structural
transition \cite{ma2}.

\subsubsection{BaFe$_2$As$_2$}

\begin{figure}
\includegraphics[width=8cm,height=8cm]{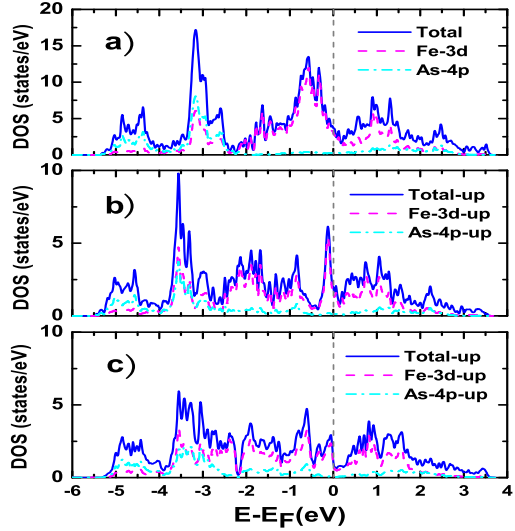}
\caption{(Color online) Total and orbital-resolved partial density
of states of BaFe$_2$As$_2$ per formula unit in a) nonmagnetic
state; b) square antiferromagnetic Neel state with spin-up (the same
for the spin-down); and c) collinear antiferromagnetic state with
spin-up (the same for the spin-down), respectively.} \label{figa}
\end{figure}

\begin{figure}
\includegraphics[width=8cm,height=8cm]{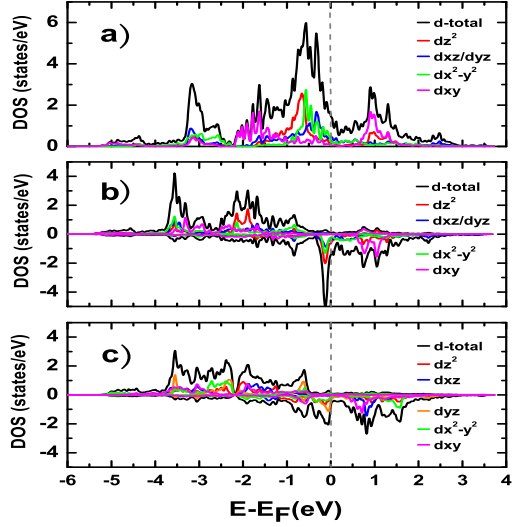}
\caption{(Color online) Total and projected density of states at the
five Fe-$3d$ orbitals of BaFe$_2$As$_2$ per Fe atom in a)
nonmagnetic state, b) square antiferromagnetic Neel state, and c)
collinear antiferromagnetic state, respectively.} \label{figb}
\end{figure}

In our calculation, the experimental tetragonal crystal lattice
parameters $a=b=3.9625 \AA$ and $c=13.0168 \AA$ \cite{rotter} were
adopted for BaFe$_2$As$_2$. Figs. \ref{figa} and \ref{figb} show the
calculated density of states (DOS) of BaFe$_2$As$_2$ in the
nonmagnetic, square antiferromagnetic Neel, and collinear
antiferromagnetic states, respectively. As revealed by Fig.
\ref{figa}(a), similar to LaFeAsO, the density of states of
BaFe$_2$As$_2$ consists of mainly the Fe-$3d$ states from -2eV to
2eV around the Fermi energy. Further analysis of the calculation
(Fig. \ref{figb}a) shows that the crystal field splitting of the
Fe-$3d$ orbitals is much smaller than the one in transition metal
oxides. This is not very surprising since the electronegativity of
As is much weaker than O. This suggests that the Fermi surface may
have the contribution from all the Fe-$3d$ orbitals.

\begin{figure}
\includegraphics[width=8cm,height=5cm]{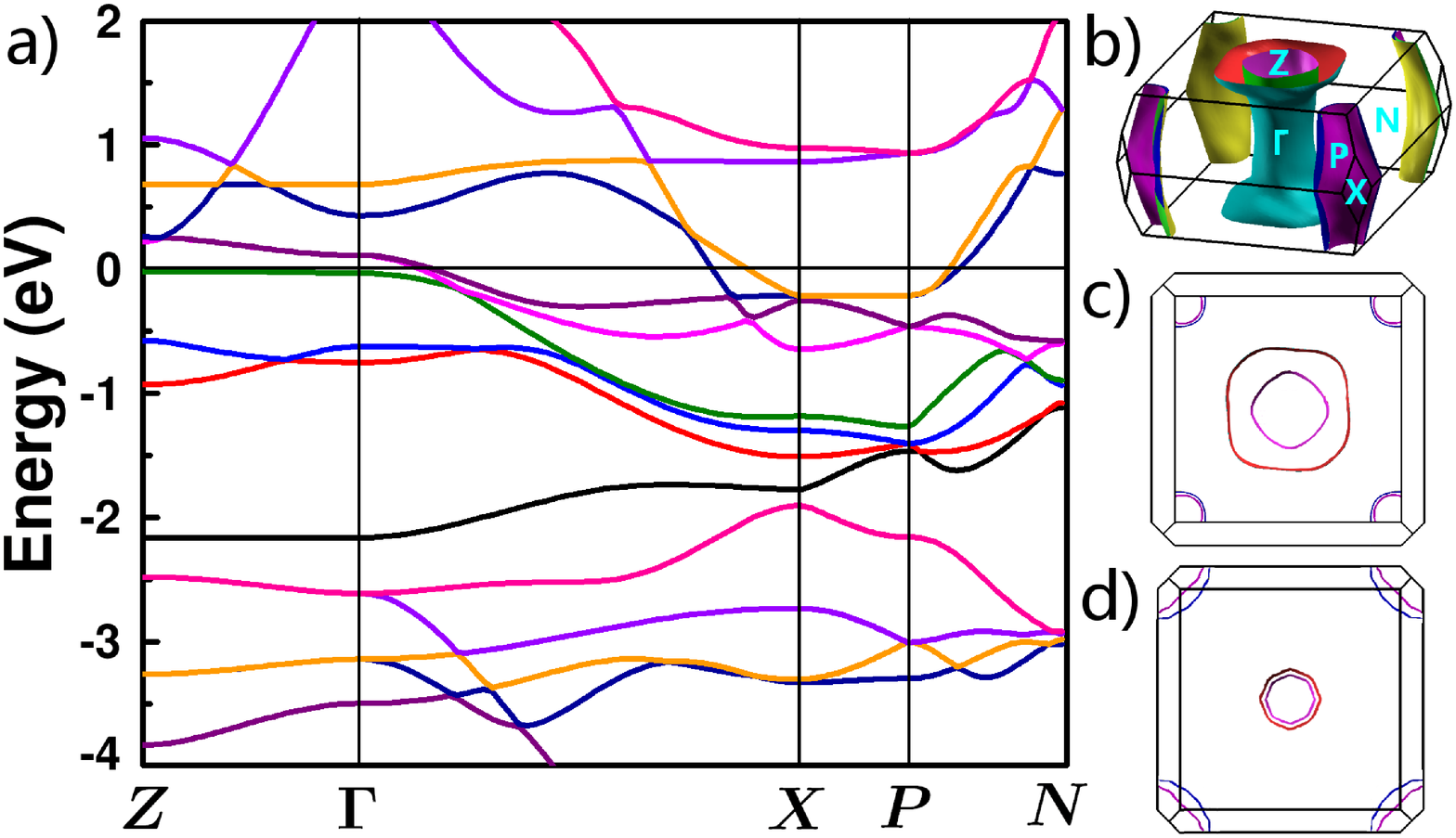}
\caption{(Color online) (a) Electronic band structure of
BaFe$_2$As$_2$ in a primitive unit cell for the nonmagnetic state.
The Fermi surface (b) and its sectional views through symmetrical
k-point $Z$ (c) and $\Gamma$ (d) parallel to (001) plane.}
\label{figc}
\end{figure}

The electronic band structure and the Fermi surface of
BaFe$_2$As$_2$ in the nonmagnetic state are shown in Fig.
\ref{figc}. As shown in Fig. \ref{figc}(a), there are four Fermi
surface sheets, contributed from the four bands crossing the Fermi
energy (Fig. \ref{figc}(a)). Among them, the two cylinder-like
Fermi-surface sheets centered around X-P are from the electron
bands. They correspond to the two Fermi surface sheets around M-A in
LaFeAsO. The other two Fermi surface sheets centered around
$\Gamma$-Z are from the hole bands. These results agree
qualitatively with the experimental observation \cite{feng}.

The energy dispersion of the electronic bands along the $c$-axis is
much larger than that in LaFeAsO\cite{singh,ma}. The sectional views
parallel to (001) plane through $Z$ point and $\Gamma$ point are
different due to the large dispersion (Figs. \ref{figc}(c) and (d)).
There is one band just below the Fermi energy along $\Gamma$-Z in
Fig. \ref{figc}(a), which corresponds to the third hole Fermi
surface sheet given in Ref. \onlinecite{nekrasov}. The volumes
enclosed by these Fermi surface sheets are 0.26 electrons/cell and
0.26 holes/cell, respectively. The electron carrier concentration is
the same as the hole carrier concentration. Both are equal to
$2.54\times 10^{21}/cm^3$. The compound BaFe$_2$As$_2$ is thus a
semimetal with a low carrier concentration between normal metals and
semiconductors, similar to what we found in LaFeAsO \cite{ma}. The
density of states at the Fermi energy is 3.93 state per eV per
formula unit cell. The corresponding electronic specific heat
coefficient $\gamma$ = $9.26mJ/(K^2\ast mol)$ and the Pauli
paramagnetic susceptibility $\chi_p$ = $1.60\times 10^{-9} m^3/mol$.
These calculated physical quantities are also summarized in Table
\ref{tableall}, which are well close to the experimental
values\cite{rotter,dong,ronning,luo,ni}.

We study the plasma excitation in the semimetal BaFe$_2$As$_2$ as
well. The plasma frequency $f_p$ is computed as follows,
\begin{eqnarray}
f_p=\frac{1}{2\pi}\sqrt{4\pi e^2({n_h\over m_h^{\ast}}+{n_e\over
m_e^{\ast}})},
\end{eqnarray}
where $n_h$ and $n_e$ are the hole and electron carrier densities,
respectively and $m_h^{\ast}$ and $m_e^{\ast}$ are the effective
masses of hole and electron, respectively. We use the second order
polynomial fitting to estimate the effective masses of carriers
from the calculated band structures. We obtain the hole and
electron effective masses as $m_e^{\ast}=0.60 m_e$ and
$m_h^{\ast}=0.86 m_e$ with $m_e$ being electron mass. Because of
irregular and strongly anisotropic band structure, the estimated
effective masses are with not small uncertainty. We then calculate
the plasma frequency as $f_p=25392\ cm^{-1}$ in the nonmagnetic
state. These values are also listed in Table \ref{tableall} in
comparison with the available experimental values.

\begin{table}
\caption{\label{tableall} Calculated (Cal.) physical quantities in
comparison with the available experimental (Exp.) values for
AFe$_2$As$_2$ (A=Ba, Sr, or Ca) in the nonmagnetic state (NM) and
the collinear antiferromagnetic state (Col). The units for the
carrier density $\rho$, the plasma frequency $f_p$, the electronic
specific heat coefficient $\gamma$, and the Pauli paramagnetic
susceptibility $\chi_{p}$ are $10^{21}/cm^{3}$, $cm^{-1}$,
$mJ/(K^{2}\ast mol)$, and $10^{-9}m^3/mol$, respectively. $J_1$ and
$J_2$ (meV/$S^2$ per Fe) are the superexchange antiferromagnetic
couplings between the nearest and the next nearest neighbor Fe spins
$\vec{S}$ in $ab$-plane, respectively while $J_z$ (meV/$S^2$ per Fe)
is the interlayer superexchange antiferromagnetic coupling of Fe
spins $\vec{S}$ perpendicular to $ab$-plane.}
    \begin{tabular}{|c|c|c|c|c|c|c|c|c|c|c|c|}
    \hline
    \multirow{2}{*}{\ AFe$_2$As$_2$\ } & \multirow{2}{*}{\ State\ } &
    \multicolumn{2}{c|}{$\rho$ } & \multicolumn{2}{c|}{$f_p$} &
     \multicolumn{2}{c|}{$\gamma$} &  \multirow{2}{*}{$\chi_p$} &
      \multicolumn{3}{c|}{Coupling} \\
    \cline{3-8}
    \cline{10-12}
    &   & hole   & electron  & Cal. & Exp. & Cal.
     & Exp. &      &   $J_1$   &   $J_2$ & $J_z$  \\
    \hline
    \multirow{2}{*}{BaFe$_2$As$_2$}  & NM  & \ 2.54 \  &\ 2.54 \
    & 25392 & 12900\cite{hu} & \ 9.26 \
     &     & \ 1.60\  & \multirow{2}{*}{\ 25.5\ } & \multirow{2}{*}{\ 33.8\ }
     & \multirow{2}{*}{\ 3.1\ } \\
    \cline{2-7}
    \cline{9-9}
    &  Col & \ 0.10 \  & \ 0.21\ & 7717 & 4660\cite{hu} & \ 5.68 \
     &  \multirow{2}{*}{
    \begin{tabular}{l}
   37\cite{ni2}, 16\cite{rotter}, \\
   6.1 \cite{dong} \\
   \end{tabular}
      }    &      &    &  &  \\
    \hline
     \multirow{2}{*}{SrFe$_2$As$_2$} &NM  & \ 3.33 \  &  \ 3.33\ & 27386
     & 13840\cite{hu} & \ 7.71 \ & \multirow{2}{*}{\ 6.5\ }  & \ 1.33\
     & \multirow{2}{*}{\ 14.7\ } & \multirow{2}{*}{\ 33.4\ }
     & \multirow{2}{*}{\ 7.8\ }\\
    \cline{2-7}
    \cline{9-9}
    &  Col & \ 0.13 \  & \ 0.04\ & 4249 & 4750\cite{hu}  & \ 3.63
     \ &     &    &      &    &  \\
    \hline
    \multirow{2}{*}{CaFe$_2$As$_2$}  &  NM  & \ 4.21 \  & \ 4.21\
     & 20883 &   & \ 9.31 \ &  \multirow{2}{*}{\ 8.2\ }   & \ 1.60\ &
     \multirow{2}{*}{\ -1.4\ } & \multirow{2}{*}{\ 26.8\ }
     & \multirow{2}{*}{\ 14.8\ }\\
    \cline{2-7}
    \cline{9-9}
    & Col & \ 0.09 \  & \ 0.09\ & 5557 &   & \ 2.62 \ &     &    &      &   &
\\
    \hline
\end{tabular}
\end{table}

\subsubsection{SrFe$_2$As$_2$}

\begin{figure}
\includegraphics[width=8cm,height=5cm]{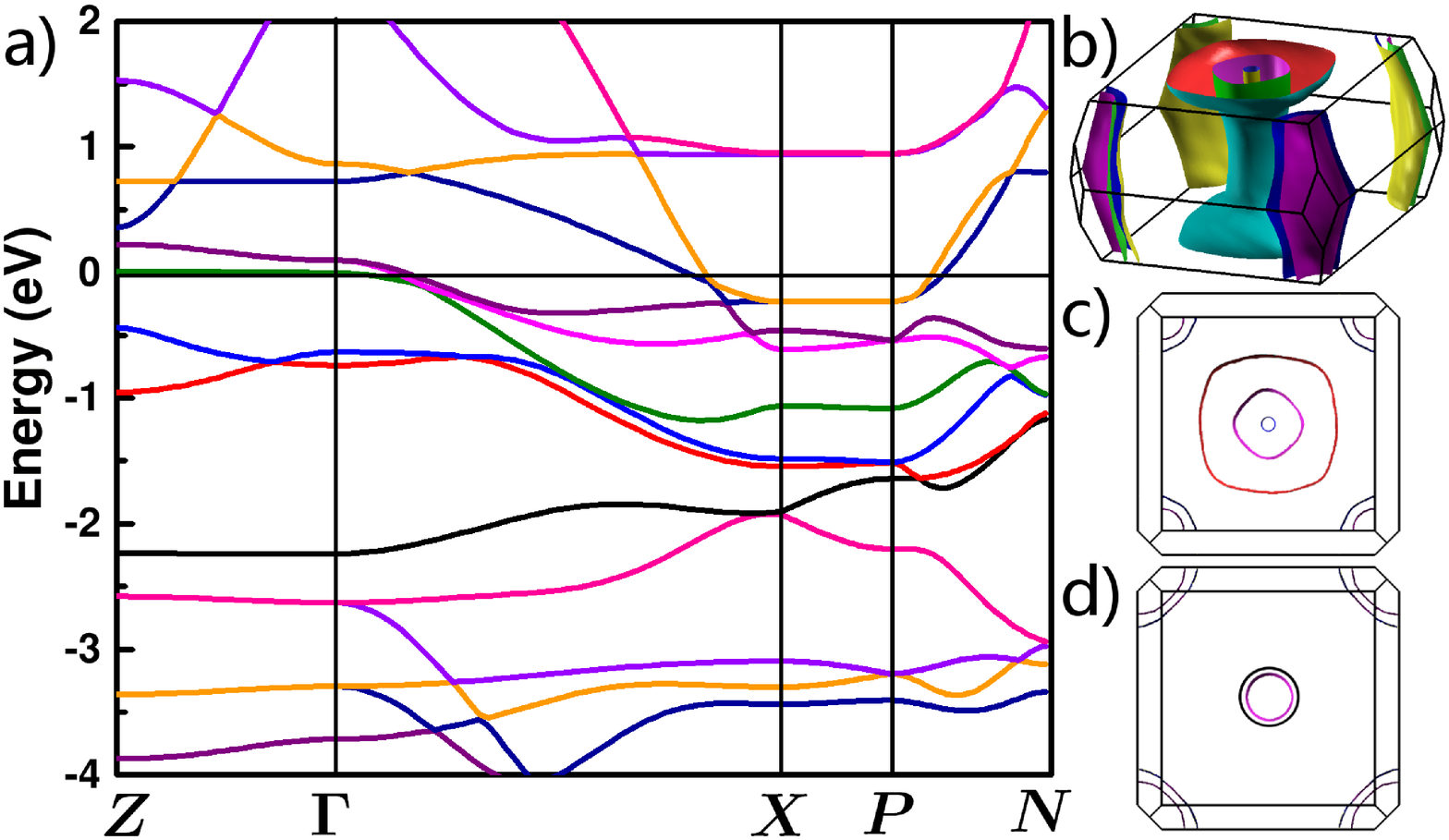}
\caption{(Color online) Calculated electronic structure of
SrFe$_2$As$_2$ in a primitive unit cell for the nonmagnetic state.
(a) Electronic band structure; The Fermi surface (b) and its
sectional views through symmetrical k-point $Z$ (c) and $\Gamma$ (d)
parallel to (001) plane. The symmetrical k-points in the Brillouin
zone are referred to Fig. \ref{figc}(b).} \label{Sr-NM-Band-Fermi}
\end{figure}

Like BaFe$_2$As$_2$, SrFe$_2$As$_2$ can also become superconducting
by chemical doping or by applying a high pressure. These two
compounds have the same crystal structure. The tetragonal crystal
lattice constants determined by experimental measurements are
$a=b=3.9259 \AA$ and $c=12.375 \AA$ for SrFe$_2$As$_2$
\cite{sasmal}. They are smaller than the corresponding parameters
for BaFe$_2$As$_2$, especially along the $c$-axis. The electronic
band structure and the Fermi surface shown in Fig.
\ref{Sr-NM-Band-Fermi}, are similar as the ones for BaFe$_2$As$_2$.
The difference is that the band along $\Gamma$-Z, which is just
below the Fermi energy in BaFe$_2$As$_2$, moves slightly upward and
intersects with the Fermi level. This results in the third hole-type
Fermi sheet centered around $\Gamma$-Z. Therefore, in
SrFe$_2$As$_2$, there are three hole-type and two electron-type
Fermi surface sheets.

The volumes enclosed by these Fermi sheets give 0.32
electrons/cell and 0.32 holes/cell for SrFe$_2$As$_2$. The
electron (or hole) carrier concentration is about $3.33\times
10^{21}/cm^3$. The density of states at the Fermi energy is about
3.27 state per eV per formula unit. The corresponding electronic
specific heat coefficient and Pauli susceptibility are $\gamma$ =
$7.71mJ/(K^2\ast mol)$ and $\chi_p$ = $1.33\times 10^{-9}
m^3/mol$, respectively. And the plasma frequency $f_p$ is computed
as about $27386\ cm^{-1}$. These calculated quantities are also
reported in Table \ref{tableall}. The total and Fe-$3d$ projected
density of states of SrFe$_2$As$_2$ are similar as the ones for
BaFe$_2$As$_2$. The low energy excitations are also dominated by
Fe-$3d$ orbitals from -2eV to 2eV around the Fermi energy.

\subsubsection{CaFe$_2$As$_2$}

\begin{figure}
\includegraphics[width=8cm,height=5cm]{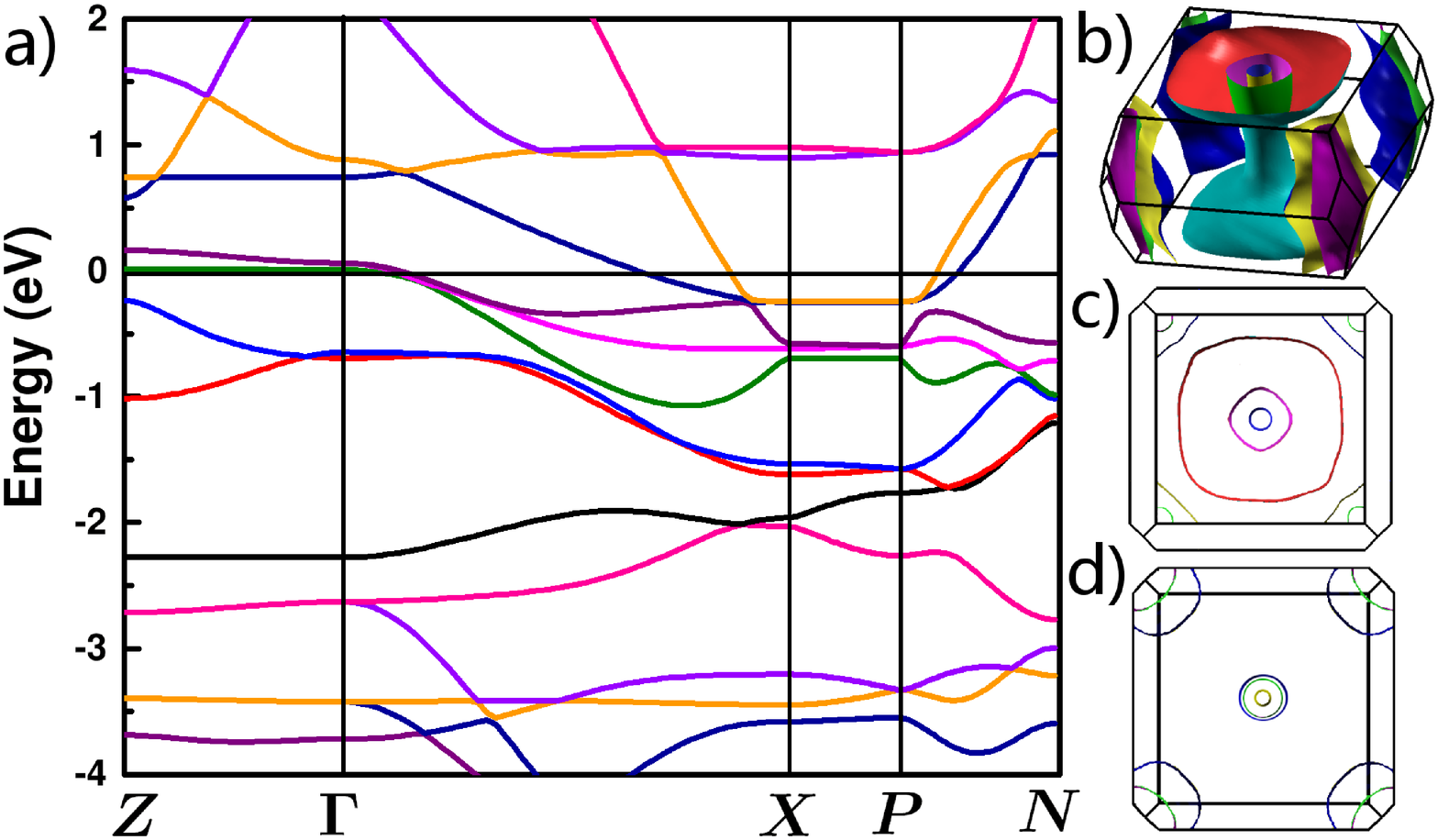}
\caption{(Color online) Calculated band structure and Fermi surface
of CaFe$_2$As$_2$ in a primitive unit cell for the nonmagnetic
state. (a) Electronic band structure; (b) Fermi surface;  And the
sectional views through symmetrical k-point $Z$ (c) and $\Gamma$ (d)
parallel to (001) plane. The symmetrical k-points in the Brillouin
zone are referred to Fig. \ref{figc}(b).} \label{Ca-NM-Band-Fermi}
\end{figure}

The tetragonal crystal lattice parameters obtained by
experiments\cite{ni}, $a=b= 3.912 \AA$ and $c=11.667 \AA$, are used
in our calculation for CaFe$_2$As$_2$. Fig. \ref{Ca-NM-Band-Fermi}
shows the electronic band structure and the Fermi surface. Similar
to SrFe$_2$As$_2$, there are three hole-type and two electron-type
Fermi surface sheets for CaFe$_2$As$_2$. Because the third hole-type
Fermi surface of CaFe$_2$As$_2$ expands into a cylinder-like shapes
centered around $\Gamma$-Z, the cross section through $\Gamma$ and X
in (001) plane has one more cutting line than that of
SrFe$_2$As$_2$. From the volumes enclosed by these Fermi sheets, we
determine the electron and hole concentrations are 0.38
electrons/cell and 0.38 holes/cell, respectively. The corresponding
electron (or hole) carrier density is about $4.21\times
10^{21}/cm^3$. The density of states at the Fermi energy is 3.95
state per eV per formula unit, and the electronic specific heat
coefficient $\gamma$ = $9.31mJ/(K^2\ast mol)$ and Pauli paramagnetic
susceptibility $\chi_p$ = $1.60\times 10^{-9} m^3/mol$. And the
plasma frequency $f_p$ is computed as about $20883\ cm^{-1}$. These
calculated quantities are summarized in Table \ref{tableall}. The
total and Fe-$3d$ orbital projected density of states of
CaFe$_2$As$_2$ are almost the same as the ones for SrFe$_2$As$_2$.

\subsection{Antiferromagnetic Neel State}

\begin{figure}
\includegraphics[width=8cm,height=5cm]{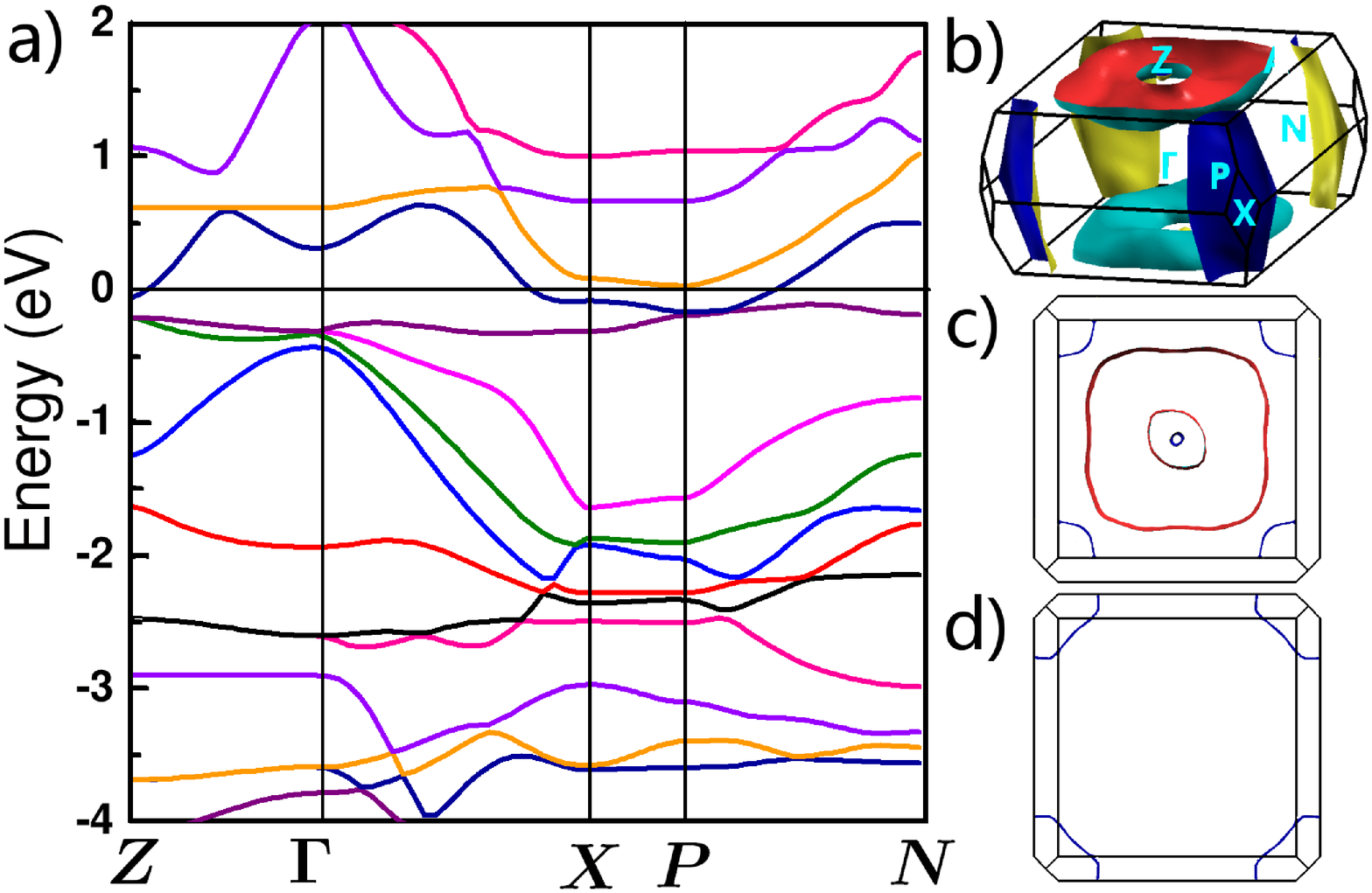}
\caption{(Color online) The electronic structure of BaFe$_2$As$_2$
in a primitive unit cell for the metastable square antiferromagnetic
Neel state. (a) Electronic band structure; The Fermi surface (b) and
its the sectional views through symmetrical k-point $Z$ (c) and
$\Gamma$ (d) parallel to (001) plane.} \label{figd}
\end{figure}

To study the electronic structures of AFe$_2$As$_2$ in a magnetic
state, we break the spin up-down symmetry by assigning a finite
magnetic moment to each Fe atom. The ferromagnetic state is found to
be not stable in AFe$_2$As$_2$, similar as in LaFeAsO \cite{ma}.
However, the antiferromagnetic Neel state is meta-stable. Its energy
is lower than the corresponding nonmagnetic state by 0.232 eV, 0.175
eV, and 0.133 eV per formula unit for BaFe$_2$As$_2$,
SrFe$_2$As$_2$, and CaFe$_2$As$_2$, respectively. The magnetic
moments in BaFe$_2$As$_2$, SrFe$_2$As$_2$, and CaFe$_2$As$_2$ are
found to be respectively 2.3 $\mu_B/Fe$, 2.2 $\mu_B/Fe$, and 2.0
$\mu_B/Fe$, similar as the one for LaFeAsO\cite{ma}, while a moment
of about 1.7 $\mu_B/Fe$ was found for all these three systems in the
full potential local-density approximation calculations
\cite{krell}. The density of states for the antiferromagnetic Neel
state of BaFe$_2$As$_2$ is shown in Figs. \ref{figa}(b) and
\ref{figb}(b). The corresponding electronic band structure with the
Fermi surface shapes is shown in Fig. \ref{figd}. The electronic
structures and the Fermi surface topology are similar for the other
two ternary iron arsenides in the antiferromagnetic Neel state.

As shown in Fig. \ref{figa}(b), the density of states around there
is substantially reduced around -0.5eV in comparison with that in
the nonmagnetic state (Fig. \ref{figa}(a)). The missing states are
pushed down to around -2.0eV. This change can be also seen by
comparing the electronic band structures of the nonmagnetic and the
square antiferromagnetic Neel states, shown in Fig. \ref{figc}(a)
and Fig. \ref{figd}(a). In contrast to the nonmagnetic state, there
are only two bands crossing the Fermi energy, forming a ring-like
hole-type shape around Z, a cylinder-like electron-type sheet
centered around X-P, and a small-pocket-like electron-type sheet
around Z.

\subsection{Collinear Antiferromagnetic State}

For BaFe$_2$As$_2$, SrFe$_2$As$_2$, and CaFe$_2$As$_2$, the
antiferromagnetic Neel state is a meta-stable state among the rich
magnetic structures. It is found that the true ground state is in
fact a collinear antiferromagnetic state with the interlayer Fe
moments in antiferromagnetic alignment. The spin configuration of
this state in FeAs layer is schematically shown in Fig.
\ref{conf}(c).

\subsubsection{BaFe$_2$As$_2$}

\begin{figure}
\includegraphics[width=11cm,height=6.5cm]{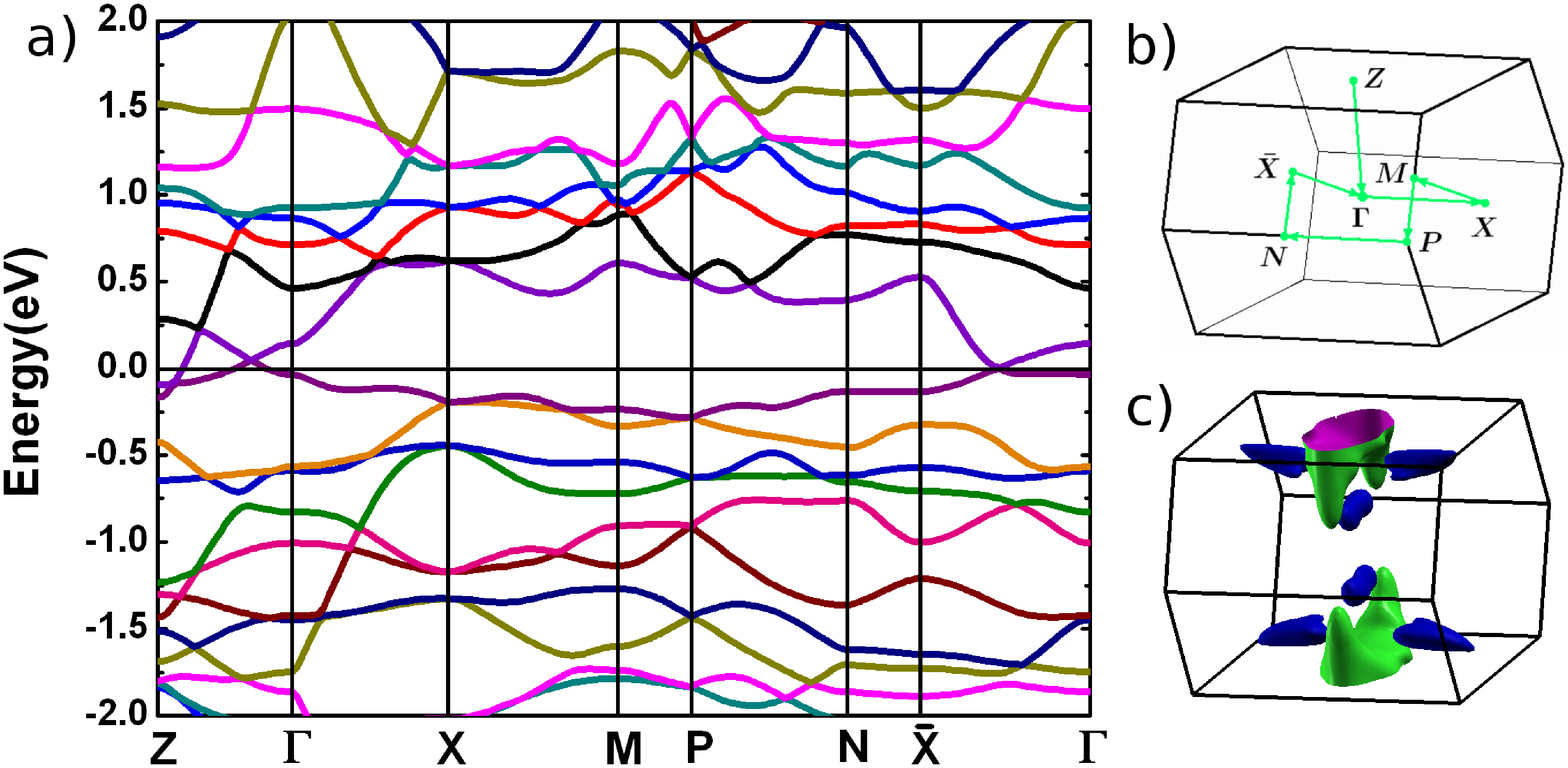}
\caption{(Color online) Calculated band structure and Fermi surface
of BaFe$_2$As$_2$ in the collinear antiferromagnetic state with the
antiparallel alignment between the interlayer Fe moments along
$c$-axis \cite{ref31}, which is the ground state. (a) Electronic
band structure; (b) Brillouin zone; (c) Fermi surface.} \label{fige}
\end{figure}

For BaFe$_2$As$_2$, the energy of the collinear antiferromagnetic
state is lowered by 0.400 eV per formula unit than that of the
nonmagnetic state. The magnetic moment is found to be about $2.65
\mu_B$ for each Fe ion in this state. The electronic band structure
and the density of states are shown in Fig. \ref{fige}(a) and Fig.
\ref{figa}(c), respectively. In contrast to the nonmagnetic state,
the density of states of As $4p$ orbitals in the low energy range
from -2eV to 0eV is substantially enhanced in the collinear
antiferromagnetic state.

Fig. \ref{fige}(c) shows the Fermi surface shapes of BaFe$_2$As$_2$
in the collinear antiferromagnetic state. There are two bands
crossing the Fermi level. They lead to a hole-type Fermi surface and
an electron-type Fermi surface, respectively. Here we remind that if
we consider BaFe$_2$As$_2$ by a tetragonal unit cell, the axis MA in
the Brillouin zone of the tetragonal unit cell is now folded into
the $\Gamma$Z, in which M point coincides with $\Gamma$ point. This
means that the electron-type Fermi sheet around M point in the
tetragonal unit cell in the nonmagnetic state will be gapped when
the collinear antiferromagnetic order takes place. From the volumes
enclosed by these two Fermi sheets, we find that the electron
carrier concentration is 0.042 electrons/cell, namely $2.05\times
10^{20}/cm^3$. The corresponding hole carrier concentration is 0.021
holes/cell, namely $1.03\times 10^{20}/cm^3$. Thus the electron
concentration dominates over the hole concentration, unlike in the
nonmagnetic state. By using the second order polynomial to fit the
calculated band structures, we estimate the electron effective
masses as $m_e^{\ast}=0.31 m_e$ with $m_e$ being electron mass. We
then calculate the plasma frequency as $f_p=7717\ cm^{-1}$. Here the
contribution of the holes to the plasma frequency is negligible.
These values are also listed in Table \ref{tableall} in comparison
with the available experimental values.

In comparison with the nonmagnetic state, the electron effective
mass becomes much lighter. Meanwhile, the carrier density is reduced
by an order of magnitude, as shown in Table \ref{tableall}. This
reduction is significantly smaller than that in LaFeAsO. In LaFeAsO,
the carrier density is reduced by two orders of magnitude in the
collinear antiferromagnetic state than in the non-magnetic state
\cite{ma2}. The density of states at the Fermi energy E$_F$ is 2.41
state per eV per formula unit cell. From this, we find that the
electronic specific heat coefficient $\gamma$ = 5.68$mJ/(K^2\ast
mol)$. It should be emphasized that the calculated value of $\gamma$
in the collinear antiferromagnetic state should be smaller than the
intrinsic value of $\gamma$ at zero temperature measured by
experiments since the low energy quantum spin fluctuations are
suppressed in the calculations, as discussed in Section \ref{sect5}.
Therefore, the measured $\gamma$ in the low temperature limit should
be bound between the two calculated $\gamma$ values obtained by our
calculations for the nonmagnetic and collinear antiferromagntic
states, respectively. In Table \ref{tableall}, we compare the
calculated specific heat coefficients $\gamma$ with the
corresponding experimental values. For BaFe$_2$As$_2$, there is a
large variety in the measurement values. Experimentally, the first
reported measurement value of $\gamma$ is about 37 $mJ/(K^2\ast
mol)$\cite{ni2}, much larger than our DFT result. However, with the
sample quality being improved, the second reported value of $\gamma$
becomes 16$mJ/(K^2\ast mol)$ \cite{rotter}; and the latest reported
value of $\gamma$ is 6.1 $mJ/(K^2\ast mol)$\cite{dong}, which is in
fact good consistent with our DFT calculation. For SrFe$_2$As$_2$
and CaFe$_2$As$_2$, our calculated results are also consistent with
the measurement values.

In the collinear antiferromagnetic state, there is further a small
energy gain if the Fe-Fe distance is reduced along the spin parallel
alignment direction and expanded along the spin anti-parallel
alignment direction. This leads to a structural transition from
tetragonal space group $I4/mmm$ to orthorhombic space group $Fmmm$,
similar to that in LaFeAsO \cite{ma2} and observed by the neutron
scattering\cite{bao}. This lattice relaxation is energetically
favorable because the direct ferromagnetic exchange favors a shorter
Fe-Fe separation while the antiferromagnetic superexchange favors a
larger Fe-As-Fe angle. It turns out that the angle $\gamma$ in
ab-plane is no longer rectangular (Fig. \ref{conf}(c)) and the
energy gain is 3 meV per formula unit, when $\gamma $ is about
$90.8^{\circ}$, similar as in LaFeAsO \cite{ma2}. The correction
from this lattice distortion to the electronic band structure as
well as the Fe moments is very small.

Along the c-axis, we find that the Fe-spins between the nearest
neighbor FeAs layers interact antiferromagnetically and the energy
gain by taking the anti-parallel alignment is about 0.012eV per
formula unit cell in comparison with the parallel alignment. This
antiferromagnetic interaction between the nearest neighboring FeAs
layers is significantly larger than that in LaFeAsO \cite{ma2}. In
Ref. \onlinecite{mafj}, we reported the electronic band structure
and the Fermi surface of AFe$_2$As$_2$ (A=Ba, Sr, Ca) in the
collinear antiferromagnetic order with the parallel alignment along
$c$-axis.

\subsubsection{SrFe$_2$As$_2$}

\begin{figure}
\includegraphics[width=11cm,height=6.5cm]{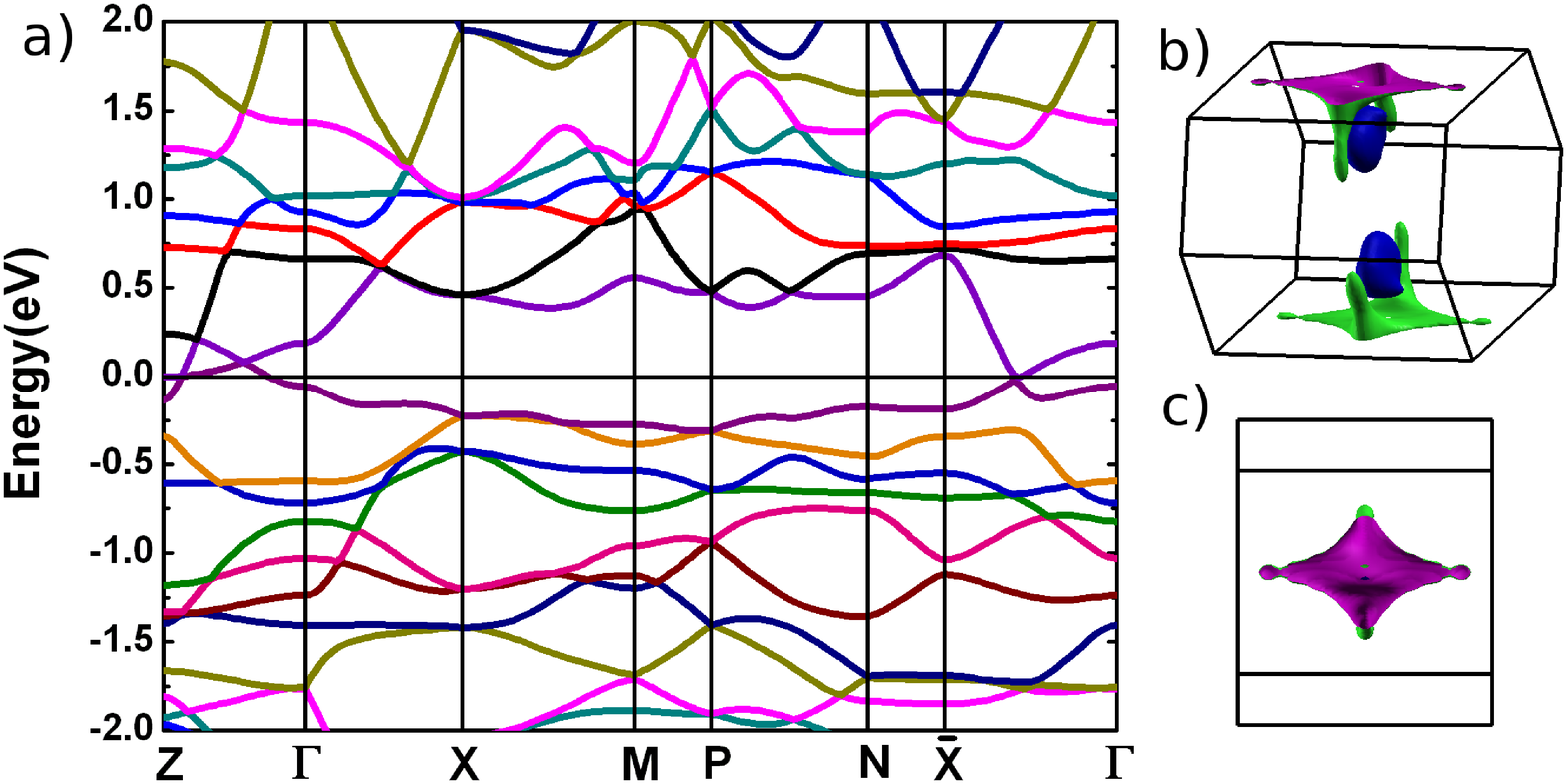}
\caption{(Color online) Calculated band structure and Fermi surface
of SrFe$_2$As$_2$ in the collinear antiferromagnetic state with the
antiparallel alignment between the interlayer Fe moments along
$c$-axis \cite{ref31}. (a) Electronic band structure; (b) Side view:
Fermi surface; (c)Top view: Fermi surface. The symmetrical k-points
in the Brillouin zone are referred to Fig. \ref{fige}(b).}
\label{Sr-Stripe-Band-Fermi}
\end{figure}

For SrFe$_2$As$_2$, the energy of the collinear antiferromagnetic
state is lowered by 0.383 eV per formula unit than the one of the
nonmagnetic state. The magnetic moment is about $2.55 \mu_B$ per Fe
atom. The electronic band structure and the Fermi surface are shown
in Fig. \ref{Sr-Stripe-Band-Fermi}. From the volumes enclosed by the
Fermi surface sheets, we find that the electron and hole carrier
concentrations are about 0.008 electrons/cell and 0.025 holes/cell,
namely about $0.44\times 10^{20}/cm^3$ and $1.31\times
10^{20}/cm^3$, respectively. It turns out that the hole
concentration dominates over the electron concentration, unlike the
case of BaFe$_2$As$_2$. And the plasma frequency $f_p$ is computed
as about $4249\ cm^{-1}$. In comparison with the nonmagnetic state,
the carrier density is much reduced, by more than an order of
magnitude. The density of states at the Fermi energy E$_F$ is 1.54
state per eV per formula unit cell, and the electronic specific heat
coefficient $\gamma$ = 3.63$mJ/(K^2\ast mol)$. These calculated
quantities are also reported in Table \ref{tableall}.

\subsubsection{CaFe$_2$As$_2$}

\begin{figure}
\includegraphics[width=11cm,height=6.5cm]{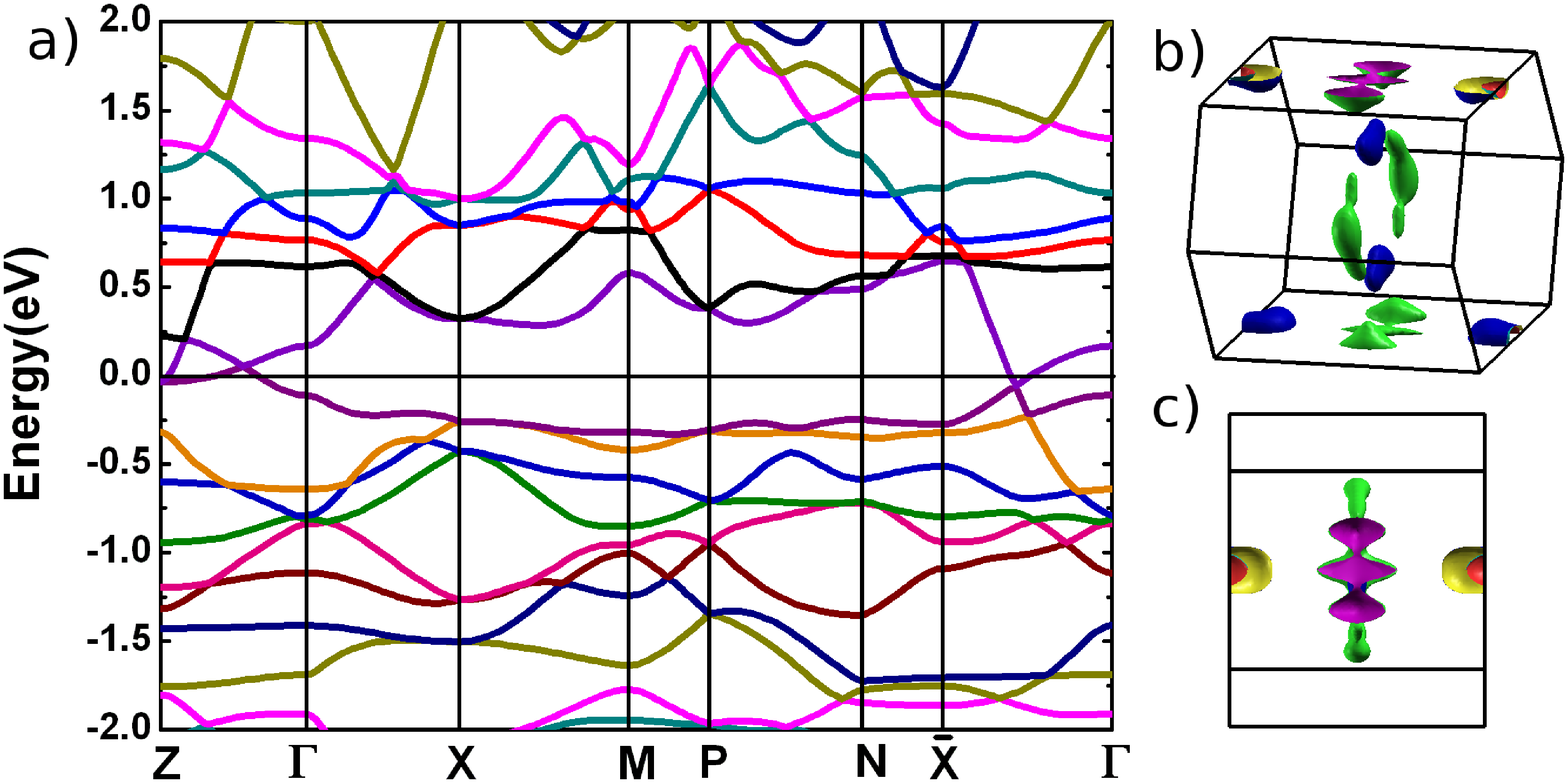}
\caption{(Color online) Calculated band structure and Fermi surface
of CaFe$_2$As$_2$ in the collinear antiferromagnetic state with the
antiparallel alignment between the interlayer Fe moments along
$c$-axis \cite{ref31}. (a) Electronic band structure; (b) Side view:
Fermi suface; (c)Top view: Fermi surface. The symmetrical k-points
in the Brillouin zone are referred to Fig. \ref{fige}(b).}
\label{Ca-Stripe-Band-Fermi}
\end{figure}

For CaFe$_2$As$_2$, the collinear antiferromagnetic ordering can
lower the ground state energy by 0.352 eV per formula unit in
comparison with the nonmagnetic state. There are about $2.49 \mu_B$
moment around each Fe atom. The electronic band structure and the
Fermi surface are shown in Fig. \ref{Ca-Stripe-Band-Fermi}. The
electron (hole) carrier density is about 0.016 electrons/cell (0.016
holes/cell), or $0.90\times 10^{20}/cm^3$ ($0.90\times
10^{20}/cm^3$). Thus the electron concentration and the hole
concentration are in balance, different from the ones in
BaFe$_2$As$_2$ and SrFe$_2$As$_2$. And the plasma frequency $f_p$ is
computed as about $5557\ cm^{-1}$. Similar to SrFe$_2$As$_2$, the
carrier density is also much reduced by more than an order of
magnitude in comparison with the nonmagnetic state. The density of
states at the Fermi energy E$_F$ is about 1.11 state per eV per
formula unit cell and the electronic specific heat coefficient
$\gamma$ = 2.62$mJ/(K^2\ast mol)$. These calculated quantities are
also reported in Table \ref{tableall}.

\subsection{Spin-Orbit Interaction}

\begin{figure}
\includegraphics[width=8cm,height=6cm]{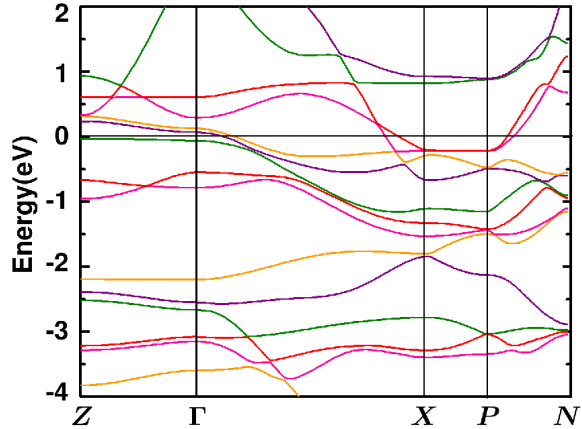}
\caption{(Color online) Electronic band structures of BaFe$_2$As$_2$
including spin-orbit interaction in the nonmagnetic
state.}\label{NM-Col-SOC}
\end{figure}

The spin-orbit interaction results from the relativistic effect. It
is known that this interaction leads to many interesting features in
transition metal oxides. To study how strong this interaction can
affect electronic properties of these materials, we performed a
relativistic calculation for BaFe$_2$As$_2$. Fig. \ref{NM-Col-SOC}
shows the electronic band structures of BaFe$_2$As$_2$ by including
spin-orbit interaction in the nonmagnetic state. By comparison with
Figs. \ref{figc}(a) and \ref{fige}(a), we find that the spin-orbit
interaction splits the band mainly along $\Gamma$-Z into two bands
by about 50$\sim$150 meV around the Fermi energy.

\subsection{Pressure Effect}

Experimentally it was reported\cite{torikachvili,park,alireza} that
the ternary iron arsenides AFe$_2$As$_2$ (A=Ba, Ca, or Sr) can
become superconducting under high pressures without doping. This
provides another route to study the salient features of these
materials in connection with the intrinsic electronic structures and
properties. The superconducting phase appears when the pressure is
in the ranges of 2.5$\sim$8, 28$\sim$37, and 22$\sim$58 kbar for
CaFe$_2$As$_2$, SrFe$_2$As$_2$, and BaFe$_2$As$_2$ respectively. The
highest superconducting transition temperatures for these three
compounds are 12K, 27K, and 29K, respectively.

\begin{figure}
\includegraphics[width=8.5cm,height=4.3cm]{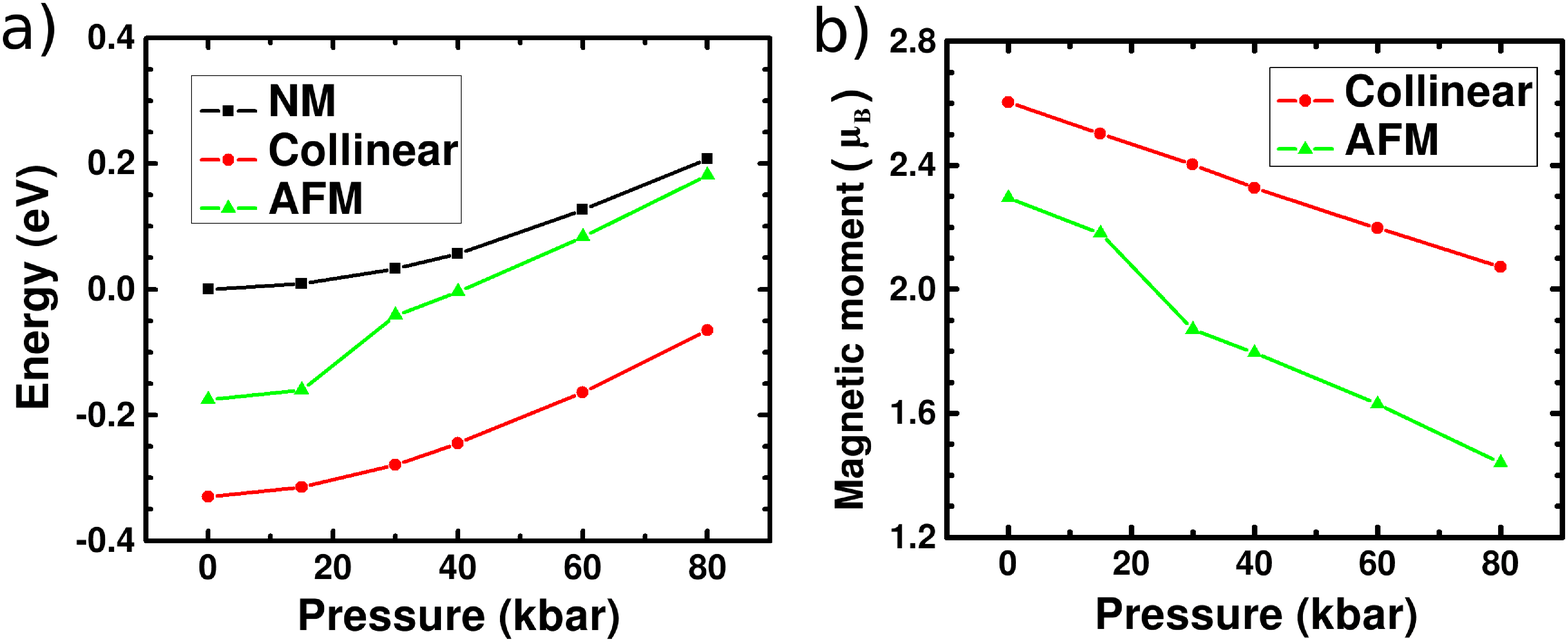}
\caption{(Color online) a) Calculated energy differences of
BaFe$_2$As$_2$ between the magnetic state and the nonmagnetic state
(NM) per formula unit cell under various pressures. Note the energy
of the nonmagnetic state without pressure is set to zero. b) The
magnetic moments for the collinear antiferromagnetic state
(Collinear) and the antiferromagnetic state (AFM) under the
pressures.} \label{VC-E-M}
\end{figure}

Fig. \ref{VC-E-M}(a) shows the pressure dependence of relative
energies for the nonmagnetic, square antiferromagnetic Neel and
collinear antiferromagnetic states of BaFe$_2$As$_2$, respectively.
The collinear antiferromagnetic state is robust against the pressure
and has the lowest energy in the pressure range studied, consistent
with the result reported in in Ref. \onlinecite{xie}. Note that in
this calculation, the lattice parameters are relaxed when the
pressure is zero. Thus the energy differences are slightly different
from that given before.

The lattice structure of FeAs layers, including the angle between Fe
and As atoms and the bond lengths of Fe-Fe, Fe-As, and As-As, is
hardly changed by the pressure from 0 to 80 kbar. However, the
contraction along the $c$ axis is much more pronounced, especially
in the nonmagnetic state. This indicates that the distance between
the neighboring FeAs and Ba layers is substantially reduced by
pressure. This can strengthen the coupling between the neighboring
FeAs layers and stabilize the long range antiferromagnetic
correlation in the FeAs layer.

The magnetic moment in the magnetic states decreases almost linearly
with pressure, as shown in Fig. \ref{VC-E-M}(b). In our
calculations, the magnetic moment is always larger than 2$\mu_B$ in
the collinear antiferromagntic state. The superexchange interaction
thus remains dominant when a pressure is imposed.

\section{Effective Model}\label{sect5}

\subsection{Local Moment versus Itinerant Electrons}

Physically the moments of Fe ions result from the on-site Coulomb
repulsion and the Hund's rule coupling of $3d$ orbitals. An isolated
Fe ion in a 2+ valency has a spin $S=2$ with a large magnetic moment
of $4 \mu_B$. In iron-pnictide semimetals, the effective moment of a
Fe ion (in a 2+ valency) will be reduced by its hybridization with
other atoms and by the Coulomb screening of itinerant electrons.
However, it will remain finite if the Hund's rule coupling and the
on-site Coulomb repulsion is strong enough in comparison with the
hybridization and other screening effects, as we found in LaFeAsO
\cite{ma2}.

In our calculations, by projecting the density of states onto the
five $3d$ orbitals of Fe in the collinear antiferromagnetic state of
BaFe$_2$As$_2$ (Fig. \ref{figb}(c)), we find that the five $3d$
orbitals of Fe are almost completely filled by up-spin electrons and
nearly half-filled by down-spin electrons in one of the two
sublattices (or completely filled by down-spin electrons and
half-filled by up-spin electrons in the other sublattice). This
indicates that the crystal field splitting imposed by As atoms is
very small and the Fe $3d$-orbitals hybridize strongly with each
other. We can see that this is a universal feature for all iron
pnictides, as we first found in LaFeAsO \cite{ma2}. The strong
polarization of Fe magnetic moments is thus due to the Hund's rule
coupling.

In low temperatures, the Fe moments will interact with each other to
form an antiferromagnetic ordered state. These ordered magnetic
moments have been observed by elastic neutron scattering and other
experiments \cite{cruz,bao}. However, they are not exactly the
moments obtained by the DFT calculations, as we indicated first for
LaFeAsO \cite{ma2}. This is because the DFT calculation is done
based on a small magnetic unit cell and the low-energy quantum spin
fluctuations as well as their interactions with itinerant electrons
are frozen by the finite excitation gap due to the finite-size
effect. Thus the moment obtained by the DFT is the bare moment of
each Fe ion. It should be larger than the ordering moment measured
by neutron scattering and other experiments.
Our calculations show that the bare magnetic moment around each Fe
atom is about $2.2\sim 2.6~\mu_B$ in all iron pnictides and in
different magnetically ordered states.

In high temperatures, there is no net static moment in the
paramagnetic phase due to the thermal fluctuation, but the bare
moment of each Fe ion can still be measured by a fast local probe
like ESR (electron spin resonance). Very recently, the bare moment
of Fe has been observed in the paramagnetic phase by the ESR
measurement \cite{chen}. The value of the moment detected by ESR is
about $2.2\sim 2.8~\mu_B$ in good agreement with our DFT result,
which is but significantly larger than the ordering moment in the
antiferromagnetic phase.

Again similar to what we found in LaFeAsO\cite{ma2}, from the
spatial distributions of electrons, we further find that there is a
strong hybridization between neighboring Fe and As ions. This strong
hybridization can mediate an antiferromagnetic superexchange
interaction between the Fe moments. This superexchange interaction
is antiferromangtic since the intermediated state associated with
the hopping bridged by As ions is a spin singlet. On the other hand,
there is a relatively small but finite hybridization between two
neighboring Fe ions. This direct hybridization of Fe $3d$ orbitals
can induce a direct exchange interaction between the Fe moments.
This direct exchange interaction is ferromagnetic due to the Hund's
rule coupling. Furthermore, there is a strong covalent bonding or
hybridization between As $4p$ orbitals although the separation
between As atoms is relatively large. This hybridization gives rise
to a broad As $4p$ band below the Fermi level. Thus As $4p$ states
are not truly localized. They form an electron network connecting As
ions through covalent bonding, similar to what we found in iron
chalcogenides ($\alpha$-FeSe and $\alpha$-FeTe)\cite{ma5}. However,
there is difference that the band formed by As $4p$-orbitals is
insulating, as shown in Fig. \ref{figa}, while the band formed by Te
$5p$-orbitals is metallic (see Ref. \onlinecite{ma5}). It turns out
that the exchange interaction bridged by As $4p$ orbitals is short
ranged (just $J_1$ and $J_2$) while the one bridged by Te $5p$
orbitals can be long ranged, as we found in $\alpha$-FeTe in which
there is a substantial third nearest neighbor antiferromagnetic
superexchange interaction $J_3$ besides $J_1$ and $J_2$ \cite{ma5}.


\subsection{$t-J_H-J_1-J_2$ Hamiltonian}

For iron-based pnictides, the low energy spin dynamics could be
approximately described by an antiferromagnetic Heisenberg model
with the nearest and the next-nearest neighbor exchange
interactions. However, the Fe spin (or magnetic moment) is not
quantized since the electrons constituting the moment can propagate
on the lattice, like in hole or electron doped high $T_c$
superconductivity cuprates. Besides these superexchange
interactions, the on-site Hund's rule coupling among different Fe
$3d$ orbitals is important. This is because the crystal splitting of
Fe $3d$ levels is very small and the spins of Fe $3d$ electrons are
polarized mainly by this interaction. Thus we believe that the
low-energy physical properties of these iron-based pnictides can be
approximately described by the following effective Hamiltonian
\begin{eqnarray}\label{eq:1}
H & = &\sum_{ \langle ij\rangle,\alpha\beta}t_{ij}^{\alpha\beta}
c_{i\alpha}^{\dagger}c_{j\beta} - J_H \sum_{i,\alpha\not=
\beta}\vec{S}_{i\alpha}\cdot\vec{S}_{i\beta}
\nonumber \\
&& + J_1\sum_{\langle ij \rangle,
\alpha\beta}\vec{S}_{i\alpha}\cdot\vec{S}_{j\beta}
+J_2\sum_{\langle\langle ij\rangle\rangle,
\alpha\beta}\vec{S}_{i\alpha} \cdot\vec{S}_{j\beta} ,
\end{eqnarray}
where $\langle ij \rangle$ and $\langle\langle ij \rangle\rangle$
represent the summation over the nearest and the next-nearest
neighbors, respectively. $\alpha$ and $\beta$ are the indices of Fe
$3d$ orbitals. $c^{\dagger}_{i\alpha}~(c_{i\alpha})$ is the electron
creation (annihilation) operator.
\begin{equation}
\vec{S}_{i\alpha} = c_{i\alpha}^\dagger \frac{\vec{\sigma}}{2}
c_{i\alpha}
\end{equation}
is the spin operator of the $\alpha$ orbital at site $i$. The total
spin operator at site $i$ is defined by
$\vec{S}_i=\sum_{\alpha}\vec{S}_{i\alpha}$. In Eq.~(\ref{eq:1}),
$J_H$ is the on-site Hund's coupling among the five Fe $3d$
orbitals. The value of $J_H$ is generally believed to be about 1 eV.
$t_{ij}^{\alpha\beta}$ are the effective hopping integrals that can
be determined from the electronic band structure in the nonmagnetic
state\cite{ccao}.

The nearest and the next-nearest neighbor antiferromagnetic coupling
constants, $J_1$ and $J_2$, in Eq.~(\ref{eq:1}) can be calculated
from the relative energies of the ferromagnetic, square
antiferromagnetic, and collinear antiferromagnetic states with
respect to the non-magnetic state. For the corresponding detailed
calculations, please refer to the appendix in our paper in Ref.
\onlinecite{ma2}. For BaFe$_2$As$_2$, the energy of the
ferromagnetic state is about 8meV per Fe lower than the nonmagnetic
state and the Fe moment is about 2.3$\mu_B$. From this and the
relative energies of the two antiferromagnetic states, we find that
the exchange constants are approximately given by $J_1$= 25.5
meV/$S^2$ per Fe and $J_2$ = 33.8 meV/$S^2$ per Fe ($S$ is the spin
of the Fe ion) for BaFe$_2$As$_2$. Meanwhile, we also find that the
interlayer superexchange antiferromagnetic coupling $J_z$ =3.1
meV/$S^2$ per Fe. In obtaining these values, we have assumed that
the contribution of itinerant electrons to the energy is almost
unchanged in different magnetically ordered states. Since the bare
$t_{ij}^{\alpha\beta}$ and $J_H$ can be considering independent of
magnetic structures, the relative energies between different
magnetic states are not affected by itinerant electrons. For
SrFe$_2$As$_2$ and CaFe$_2$As$_2$, the values of $J_1$ and $J_2$
with $J_z$ are determined as well and given in Table \ref{tableall}.
As we notice, $J_1$ in CaFe$_2$As$_2$ is negative, namely
ferromagnetic, which is unique among iron pnictides.

\subsection{Discussion}

\begin{figure}
\includegraphics[width=7.5cm,height=7.5cm]{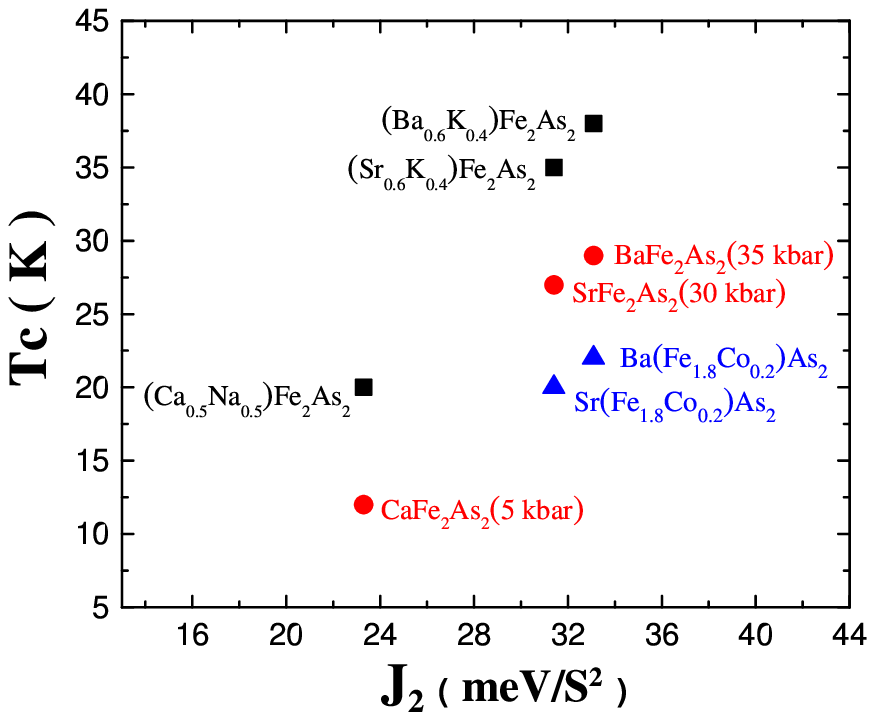}
\caption{Various critical superconductivity temperature $T_c$ versus
exchange interaction $J_2$ for AFe$_2$As$_2$ (A=Ba, Ca, or Sr) with
doping or under high pressures. Tc are taken from Ref.
\onlinecite{rotter2,wu,luo,alireza,leithe,sefat}.}\label{Tc-J2}
\end{figure}

\begin{figure}
\includegraphics[width=7.5cm,height=5cm]{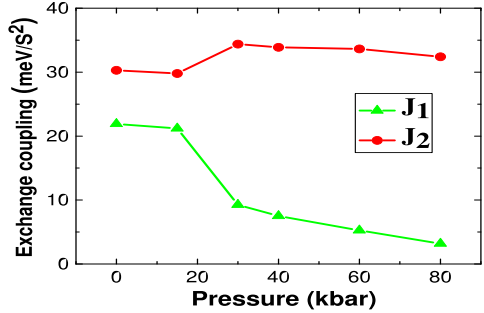}
\caption{Variations of superexchange antiferromagnetic interaction
constants $J_1$ and $J_2$ with the imposed pressure for
BaFe$_2$As$_2$. }\label{VC-J1-J2}
\end{figure}

We plot the superconducting critical temperatures $T_c$ versus $J_2$
for these compounds with doping carriers or by applying high
pressures in Fig. \ref{Tc-J2}. Interesting, as we see, the maximum
critical temperatures $T_c$ are in proportion to the next-nearest
neighbor superexchange interaction $J_2$. This suggests that there
would exist an intrinsic relationship between the superconductivity
and the As-bridged superexchange antiferromagnetic interactions.
Here we have assumed that the value of $J_2$ does not change with
doping or pressure in comparison with the ones of the parent
compounds. This can be verified from Fig. \ref{VC-J1-J2}. Fig.
\ref{VC-J1-J2} shows how the superexchange interactions $J_1$ and
$J_2$ change with the pressure, from which we find that $J_2$
slightly changes with the pressure. However, $J_1$ drops quickly
with increasing pressure. This is because the energy difference
between the non-magnetic state and the square antiferromagnetic Neel
state can be significantly reduced by the pressure, as shown in Fig.
\ref{VC-E-M}.

In comparison with the bare $t_{ij}^{\alpha\beta}$ and $J_H$, $J_1$
and $J_2$ are very small. When we study charge dynamics like charge
transport happening in iron pnictides, $t_{ij}^{\alpha\beta}$-$J_H$
part will thus play a dominant role. In Eq.~(\ref{eq:1}), if the
$J_1$ and $J_2$ terms are ignore, the Hamiltonian will look similar
to the double exchange model that was proposed for describing
physical properties of colossal magneto-resistance (CMR) of
manganese oxides\cite{von,zener}. However, there is an essential
difference. In manganese oxides the crystal field splitting between
$e_g$ and $t_{2g}$ orbitals is very large, the moments polarized by
the Hund's rule coupling on the $t_{2g}$ orbitals are completely
localized and only $e_g$ electrons can hop on the lattice. In
contrast, in iron pnictides, all five $3d$ orbitals of Fe have
contributions to the moment and at the same time they can also hop
on the lattice. Nevertheless, we believe that their charge dynamics
shares a common feature. In manganese oxides there is a strong
spin-dependent scattering on the conduction electrons due to the
strong Hund's coupling between $e_g$ and $t_{2g}$ levels of Mn ions.
If the localized core spins of Mn are aligned ferromagnetically, the
scattering of the electrons due to the core spins will be
dramatically reduced because of no spin-flip, leading to a giant
magneto-resistance and a sharp drop of resistivity with decreasing
temperature in the ferromagnetic phase. In iron pnictides, there is
also a strong spin-dependent scattering on conduction electrons
caused by the Hund's rule coupling. In low temperatures, the moments
of Fe ions are in the collinear antiferromagnetic order, in which
the Fe moments are aligned antiferromagnetically along one direction
but ferromagnetically along the other direction perpendicular.
Similar as in manganese oxides, the scattering of electrons along
the ferromagnetic direction is significantly reduced, leading to a
sharp drop of resistivity and a large magneto-resistivity in the
collinear antiferromagnetic phase, in agreement with experimental
measurements\cite{luo,cheng}.

\section{Conclusion}

In conclusion, we have reported calculated results on the electronic
band structures of AFe$_2$As$_2$ (A=Ba, Ca, or Sr) by using the
first-principles electronic structure calculations. The ground state
of AFe$_2$As$_2$ is shown to be a collinear antiferromagnetic
semimetal with a large magnetic moment around each Fe ion. The
electronic structure is weakly affected by the spin-orbit
interaction, but strongly altered by pressure. We have determined
the density of states at the Fermi level, the specific heat
coefficient and the Pauli susceptibility in both the non-magnetic
and collinear antiferromagnetic states. The effective
antiferromagnetic coupling constants of the Fe moments are also
estimated assuming that the low-energy spin dynamics is
approximately described by the Heisenberg model with the nearest and
the next-nearest neighboring exchange terms. Based on the analysis
of electronic and magnetic structures, we proposed that the
low-energy physics of AFe$_2$As$_2$ can be effectively described by
the $t-J_H-J_1-J_2$ model, defined by Eq.~(\ref{eq:1}).


This work is partially supported by National Natural Science
Foundation of China and by National Program for Basic Research of
MOST, China.


\begin{references}

\bibitem{mafj} Part of the calculations presented in this paper had been first
reported in our paper arXiv:0806.3526v2

\bibitem{kamihara}
Kamihara Y, Watanabe T, Hirano M and Hosono H 2008
\newblock Iron-Based Layered Superconductor La[O$_{1-x}$F$_x$]FeAs (x =
0.05-0.12) with $T_c$ = 26 K {\em J. Am. Chem. Soc.} {\bf 130} 3296

\bibitem{rotter2}
Rotter M, Tegel M and Johrendt D 2008
\newblock Superconductivity at 38 K in the Iron Arsenide
(Ba$_{1-x}$K$_{x}$)Fe$_2$As$_2$ {\em Phys. Rev. Lett.} {\bf 101}
107006

\bibitem{sasmal}
Sasmal K, Lv B, Lorenz B, Guloy A, Chen F, Xue Y and  Chu C W 2008
\newblock Superconducting Fe-Based Compounds (A$_{1-x}$Sr$_x$)Fe$_2$As$_2$ with
A=K and Cs with Transition Temperatures up to 37 K {\em Phys. Rev.
Lett.} {\bf 101} 107007

\bibitem{wu}
Wu G, Chen H, Wu T, Xie Y L, Yan Y J, Liu R H, Wang X F, Ying J J
and Chen H H 2008
\newblock Different resistivity response to spin density wave and
superconductivity at 20 K in Ca$_{1-x}$Na$_x$Fe$_2$As$_2$ {\em
Journal of Physics: Condensed Matter } {\bf 20} 422201

\bibitem{boyer}
Boyer M C, Chatterjee K, Wise W D, Chen G F, Luo J L, Wang N L and
Hudson E W 2008
\newblock Scanning tunneling microscopy of the 32 K superconductor
(Sr$_{1-x}$K$_x$)Fe$_2$As$_2$ \textit{Preprint} arXiv:0806.4400

\bibitem{cruz}
Cruz C de la, Huang Q, Lynn J W, Li J, Ratcliff W, Zarestky J L,
Mook H A, Chen G F, Luo J L, Wang N L and Dai P 2008
\newblock Magnetic order close to superconductivity in the iron-based Layered
La(O$_{1-x}$F$_x$)FeAs systems {\em Nature} {\bf 453} 899

\bibitem{mcguire}
McGuire M A, Christianson A D, Sefat A S, Jin R, Payzant E A, Sales
B C, Lumsden M D and Mandrus D 2008
\newblock Phase transitions in LaFeAsO: Structural, magnetic, elastic, and
transport properties, heat capacity and M\"{o}ssbauer spectra {\em
Phys. Rev. B} {\bf 78} 094517

\bibitem{rotter}
Rotter M, Tegel M, Schellenberg I, Hermes W, P\"{o}ttgen R and
Johrendt D 2008
\newblock Spin-density-wave anomaly at 140 K in the ternary iron arsenide
BaFe$_2$As$_2$ {\em Phys. Rev. B} {\bf 78} 020503(R)

\bibitem{bao}
Huang Q, Qiu Y, Bao W, Lynn J W, Green M A, Gasparovic Y C, Wu T, Wu
G and Chen X H 2008
\newblock Neutron-Diffraction Measurements of Magnetic Order and a Structural
Transition in the Parent BaFe$_2$As$_2$ Compound of FeAs-Based
High-Temperature Superconductors {\em Phys. Rev. Lett.} {\bf 101}
257003

\bibitem{torikachvili}
Torikachvili M S, Bud'ko S L, Ni N and Canfield P C 2008
\newblock Pressure Induced Superconductivity in CaFe$_2$As$_2$ {\em Phys. Rev.
Lett.} {\bf 101} 057006

\bibitem{park}
Park T, Park E, Lee H, Klimczuk T, Bauer E D, Ronning F and Thompson
J D 2008
\newblock Pressure-induced superconductivity in CaFe$_2$As$_2$ {\em J. Phys.:
Condens. Matter} {\bf 20} 322204

\bibitem{alireza}
Alireza P L, Gillett J, Ko Y T C, Sebastian S E and Lonzarich G G
2009
\newblock Superconductivity up to 29 K in SrFe$_2$As$_2$ and BaFe$_2$As$_2$ at
high pressures {\em J. Phys.: Condens. Matter} {\bf 21} 012208

\bibitem{nekrasov}
Nekrasov I A, Pchelkina Z V and Sadovskii M V 2008
\newblock Electronic Structure of Prototype AFe$_2$As$_2$ and ReOFeAs
High-Temperature Superconductors: a Comparison {\em JETP Letters}
{\bf 88} 144

\bibitem{singh2}
Singh D J 2008
\newblock Electronic structure and doping in BaFe$_2$As$_2$ and LiFeAs: Density
functional calculations {\em Phys. Rev. B} {\bf 78} 094511

\bibitem{pwscf}
Giannozzi P et al.
\newblock {\em http://www.quantum-espresso.org}

\bibitem{pbe}
Perdew J P, Burke K and Ernzerhof M 1996
\newblock Generalized Gradient Approximation Made Simple {\em Phys. Rev. Lett.}
{\bf 77} 3865

\bibitem{vanderbilt}
Vanderbilt D 1990
\newblock Soft self-consistent pseudopotentials in a generalized eigenvalue
formalism {\em Phys. Rev. B} {\bf 41} 7892

\bibitem{ma2}
Ma F, Lu Z Y and Xiang T 2008
\newblock Arsenic-bridged antiferromagnetic superexchange interactions in
LaFeAsO {\em Phys. Rev. B} {\bf 78} 224517

\bibitem{feng}
Yang L X, Zhang Y, Ou H W, Zhao J F, Shen D W, Zhou B, Wei J, Chen
F, Xu M, He C, Chen Y, Wang Z D, Wang X F, Wu T, Wu G, Chen X H,
Arita M, Shimada K, Taniguchi M, Lu Z Y, Xiang T and Feng D L 2009
\newblock Electronic Structure and Unusual Exchange Splitting in the
Spin-Density-Wave State of the BaFe$_2$As$_2$ Parent Compound of
Iron-Based Superconductors {\em Phys. Rev. Lett.} {\bf 102} 107002

\bibitem{singh}
Singh D J and Du M H 2008
\newblock Density Functional Study of LaFeAsO$_{1-x}$F$_x$: A Low Carrier
Density Superconductor Near Itinerant Magnetism {\em Phys. Rev.
Lett.} {\bf 100} 237003

\bibitem{ma}
Ma F and Lu Z Y 2008
\newblock Iron-based layered compound LaFeAsO is an antiferromagnetic semimetal
{\em Phys. Rev. B} {\bf 78} 033111

\bibitem{dong}
Dong J K, Ding L, Wang H, Wang X F, Wu T, Wu G, Chen X H and Li S Y
2008
\newblock Thermodynamic properties of Ba$_{1-x}$K$_x$Fe$_2$As$_2$ and
Ca$_{1-x}$Na$_x$Fe$_2$As$_2$ {\em New Journal of Physics} {\bf 10}
123031

\bibitem{ronning}
Ronning F, Klimczuk T, Bauer E D, Volz H and Thompson J D 2008
\newblock Synthesis and properties of CaFe$_2$As$_2$ single crystals {\em J.
Phys.: Condens. Matter} {\bf 20} 322201

\bibitem{luo}
Chen G F, Li Z, Dong J, Li G, Hu W Z, Zhang X D, Song X H, Zheng P,
Wang N L and Luo J L 2008
\newblock Transport and anisotropy in single-crystalline SrFe$_2$As$_2$ and
A$_{0.6}$K$_{0.4}$Fe$_2$As$_2$ (A=Sr, Ba) superconductors {\em Phys.
Rev. B} {\bf 78} 224512

\bibitem{ni}
Ni N, Nandi S, Kreyssig A, Goldman A I, Mun E D, Bud'ko S L and
Canfield P C 2008
\newblock First-order structural phase transition in CaFe$_2$As$_2$ {\em Phys.
Rev. B} {\bf 78} 014523

\bibitem{hu}
Hu W Z, Dong J, Li G, Li Z, Zheng P, Chen G F, Luo J L and Wang N L
2008
\newblock Origin of the Spin Density Wave Instability in AFe$_2$As$_2$ (A=Ba,Sr)
as Revealed by Optical Spectroscopy {\em Phys. Rev. Lett.} {\bf 101}
257005

\bibitem{ni2}
Ni N, Bud'ko S L, Kreyssig A, Nandi S, Rustan G E, Goldman A I,
Gupta S, Corbett J D, Kracher A and Canfield P C 2008
\newblock Anisotropic thermodynamic and transport properties of
single-crystalline Ba$_{1-x}$K$_x$Fe$_2$As$_2$ (x=0 and 0.45) {\em
Phys. Rev. B} {\bf 78} 014507

\bibitem{krell}
Krellner C, Caroca-Canales N, Jesche A, Rosner H, Ormeci A and
Geibel C 2008
\newblock Magnetic and structural transitions in layered iron arsenide systems:
AFe$_2$As$_2$ versus RFeAsO {\em Phys. Rev. B} {\bf 78} 100504

\bibitem{ref31}
In Ref. \onlinecite{mafj}, we had reported the electronic band
structure and the Fermi surface of AFe$_2$As$_2$ (A=Ba, Sr, Ca) in
the collinear antiferromagnetic order with the parallel alignment
between interlayer Fe moments along $c$-axis.

\bibitem{xie}
Xie W H, Bao M L, Zhao Z J and Liu B G 2009
\newblock First-principles investigation of the effect of pressure on
BaFe$_2$As$_2$ {\em Phys. Rev. B} {\bf 79} 115128

\bibitem{ma5}
Ma F, Ji W, Hu J P, Lu Z Y and Xiang T 2009
\newblock First-Principles Calculations of the Electronic Structure of
Tetragonal $\alpha$-FeTe and $\alpha$-FeSe Crystals: Evidence for a
Bicollinear Antiferromagnetic Order {\em Phys. Rev. Lett.} {\bf 102}
177003

\bibitem{ccao}
Cao C, Hirschfeld P J and Cheng H P 2008
\newblock Proximity of antiferromagnetism and superconductivity in
LaFeAsO$_{1-x}$F$_x$: Effective Hamiltonian from ab initio studies
{\em Phys. Rev. B} {\bf 77} 220506(R)

\bibitem{leithe}
Leithe-Jasper A, Schnelle W, Geibel C and Rosner H
\newblock Superconductivity in SrFe$_{2-x}$Co$_x$As$_2$: Internal Doping of the
Iron Arsenide Layers \textit{Preprint} arXiv:0807.2223

\bibitem{sefat}
Sefat A S, Jin R, McGuire M A, Sales B C, Singh D J and Mandrus D
2008
\newblock Superconductivity at 22 K in Co-Doped BaFe$_2$As$_2$ Crystals {\em
Phys. Rev. Lett.} {\bf 101} 117004

\bibitem{von}
Helmolt R V, Wocker J, Holzapfel B, Schultz L and Samwer K 1993
\newblock Giant negative magnetoresistance in perovskitelike
La$_{2/3}$Ba$_{1/3}$MnO$_x$ ferromagnetic films {\em Phys. Rev.
Lett.} {\bf 71} 2331

\bibitem{zener}
Zener C 1951
\newblock Interaction between the d-Shells in the Transition Metals. II.
Ferromagnetic Compounds of Manganese with Perovskite Structure {\em
Phys. Rev.} {\bf 82} 403

\bibitem{cheng}
Cheng P, Yang H, Jia Y, Fang L, Zhu X, Mu G and Wen H H 2008
\newblock Hall effect and magnetoresistance in single crystals of
NdFeAsO$_{1-x}$F$_x$ (x=0 and 0.18) {\em Phys. Rev. B} {\bf 78}
134508

\bibitem{gfchen}
Chen G F, Li Z, Dong J, Li G, Hu W Z, Zhang X D, Song X H, Zheng P,
Wang N L and Luo J L 2008
\newblock Transport and anisotropy in single-crystalline SrFe$_2$As$_2$ and
A$_{0.6}$K$_{0.4}$Fe$_2$As$_2$ (A=Sr, Ba) superconductors {\em Phys.
Rev. B} {\bf 78} 224512

\bibitem{chen}
Wu T, Ying J J, Wu G, Liu R H, He Y, Chen H, Wang X F, Xie Y L, Yan
Y J and Chen X H 2009
\newblock Evidence for local moments by electron spin resonance study of
polycrystalline LaFeAsO$_{1-x}$F$_x$ (x=0 and 0.13) {\em Phys. Rev.
B} {\bf 79} 115121


\end{references}
\end{document}